\documentclass[twocolumn,aps,prb,superscriptaddress]{revtex4}
\usepackage{graphics,bm,graphicx}
\graphicspath{{Users/samolin/Downloads/TMD-GK-Competition/}}
\usepackage{amssymb}
\usepackage{epsfig}
\usepackage{epsf}
\usepackage{xcolor}
\usepackage{tabularx}

\def\be{\begin{equation}} \def\ee{\end{equation}}
\def\beq{\begin{eqnarray}} \def\eeq{\end{eqnarray}}

\newcommand{\angstrom}{\text{\normalfont\AA}}

\begin{document}

\title{Ab-initio study of the energy competition between \\ $\Gamma$ and K valleys in bilayer transition metal dichalcogenides}

\author{Sam Olin}
\email{solin1@binghamton.edu}
\affiliation{Department of Physics, Applied Physics, and Astronomy, Binghamton University, Binghamton, New York, 13902, USA}

\author{Erekle Jmukhadze}
\affiliation{Department of Physics, Applied Physics, and Astronomy, Binghamton University, Binghamton, New York, 13902, USA}

\author{Allan H. MacDonald}
\affiliation{Department of Physics, the University of Texas at Austin, Austin, Texas 78712, USA}

\author{Wei-Cheng Lee}
\email{wlee@binghamton.edu}
\affiliation{Department of Physics, Applied Physics, and Astronomy, Binghamton University, Binghamton, New York, 13902, USA}

\date{\today}

\begin{abstract}
Moir\'e engineering in two-dimensional van der Waals bilayer crystals has emerged as a flexible platform for controlling  
strongly correlated electron systems. 
The competition between valleys for the band extremum energy position 
in the parent layers is crucial in deciding the qualitative nature of the 
moir\'e Hamiltonian since it controls the physics of the moir\'e minibands.  
Here we use density functional theory to examine the competition between K and $\Gamma$ for the valence band maximum 
in homo- and hetero-bilayers formed from the transition metal dichalcogenides (TMD),
$\textrm{MX}_2$ where $\textrm{M}=\textrm{Mo},\textrm{W}$ and $\textrm{X}=\textrm{S},\textrm{Se},\textrm{Te}$.
We shed light on how the competition is influenced by interlayer separation, which can be modified by 
applying pressure, by external gate-defined electric fields, and by
transition metal atom $d$-orbital correlations. Our findings are related to several recent experiments, 
and contribute to the development of design rules for moir\'{e} materials.
\end{abstract}

\maketitle

\section{Introduction}
Because their miniband widths can be narrow, 
two-dimensional van der Waals material multilayer moir\'es provide 
an attractive platform for designer quantum materials. 
One example of moir\'{e} engineering is provided by twisted bilayer graphene, in which
novel strongly correlated electronic states, including Mott insulators and superconductors, emerge only at certain magic angles between layers\cite{Cao_2018-insulator,Cao_2018-supcon,Lu_2019,PhysRevLett.121.026402,PhysRevLett.121.257001,PhysRevX.8.031089}. The basic mechanism of moir\'{e} engineering is periodicity at a length scale that is controllable, and in a range that allows the number of electrons per effective atom to be tuned over ranges larger then one using electrical gates.
In twisted bilayer graphene flat bands emerge when the interference between intralayer and interlayer hopping is destructive. \cite{Bistritzer_2011, PhysRevB.81.161405}

 
Recently group VI transition metal dichalcogenides (TMDs) have stimulated enormous research interest
because of their potential to host even richer physics under moir\'{e} engineering \cite{PhysRevX.8.031089,PhysRevLett.122.016401,PhysRevLett.121.266401,PMID:32572205,moire-mott,Jin_2019}. 
In monolayer form, these materials are semiconductors with direct bandgaps in which both conduction band minimum (CBM) and valence band maximum (VBM) are at the K point \cite{C4CS00265B,Li-bandgrapscreen} in the Brillouin zone.
Non-trivial Berry curvature near the K points gives rise to a number of unique electronic and optical properties.\cite{xiao2012, Mak2018} For bilayer TMDs, the VBM can be at $\Gamma$ or K points depending on a variety of factors \cite{Angeli_2021,doi:10.1021/acs.jpcc.6b08748,doi:10.1021/ar5002846,https://doi.org/10.1002/andp.201400128,PhysRevB.86.075454,doi:10.1021/jp075424v,doi:10.1021/nl302584w}. Since the location of the VBM controls
optical properties, the moir\'{e} potential landscape, and the topological properties of the moir\'{e} bandstructures, 
a systematic investigation of the key factors that determine the location of the VBM in bilayer TMDs is timely \cite{doi:10.1021/acsomega.0c01138,Angeli_2021,Koperski,Kumar_2013,doi:10.1021/acs.chemmater.7b00453,PhysRevX.12.021065,Kozawa_2014}. 

In this paper, we employ first-principles methods to examine the trends in position of the VBM in twist-free homo- and hetero-bilayer TMDs, demonstrating that it is determined by a competition between interlayer tunneling, spin-orbit coupling, applied gate voltage, and electron correlation influences on the $d$ orbitals of the transition metal atoms.  In particular, stronger interlayer tunneling favors the $\Gamma$ point while stronger spin-orbit coupling favors the K point. On top of these two opposing factors, the influence of electron correlations on the $d$ orbitals tends to increase energy in states with more localized wave functions, and therefore to raise the energy of states near the K point.  In this paper we first explore these competitions at greater depth and then discuss experimentally feasible strategies to control the VBM location for moir\'{e} engineering in light of our findings.


\section{Methods}
\subsection{Structural Relaxation}\label{section:Struct-relax}
Herein we consider twist-free, homo- and hetero- bilayer systems with chemical formula $\textrm{MX}_2$ where $\textrm{M}=\textrm{Mo},\textrm{W}$ and $\textrm{X}=\textrm{S},\textrm{Se},\textrm{Te}$. For heterobilayers, we consider only combinations containing common chalcogens in the parent layers to avoid possible inaccuracies due to large lattice constant mismatches. While many bilayer stacking arrangements exist, in this work we consider only the two high symmetry stacking orders denoted as 2H and AA and displayed in Fig. \ref{fig:stacking-def}. The bilayer AA stacking occurs when the $\hat{x},\hat{y}$ coordinates of the metal (chalcogen) atom(s) in the top layer are the same as the metal (chalcogen) atom(s) in the bottom layer \cite{Devakul_2021}. The 2H stacking is obtained by a $180^{\circ}$ rotation about the $\hat{z}$-axis of the top layer relative to the AA stacking case, and is predicted to be the most stable of the bilayer configurations \cite{doi:10.1126/sciadv.ade5706,PhysRevB.89.075409}.
In this case, the metal (chalcogen) atom(s) are directly above the chalcogen (metal) atom(s). These arrangements can not be made equal by relative translations of the layers and therefore do not occur at different positions in the same moir\'e pattern.

We employ the Vienna ab-initio Simulation Package (VASP)\cite{PhysRevB.47.558,KRESSE199615,PhysRevB.54.11169} to perform structural relaxation for each 2H system under the following procedure. Firstly, full relaxation in which both the volume and shape of the unit cell may vary to minimize the total energy and force is performed on the bulk structures using three functionals; the Perdew-Burke-Ernzerhof exchange-correlation functional with spin orbit coupling, both with (PBE-SO-D3) and without (PBE-SO) the van der Waals correction, and the local density approximation LDA+SO. Additionally, we include the relaxation via PBE alone (as performed in [\onlinecite{materials-project}]) for comparison. Secondly, we construct each 'free-standing' bilayer system from the bulk lattice constants found from the LDA+SO relaxation. The total length of the c-axis is set to 35\AA (Fig. \ref{fig:stacking-def}), to isolate the bilayer and limit the unphysical interaction between periodic unit cells. This structure is then relaxed with fixed volume using LDA-SO in VASP, allowing atomic positions to change. From these structures, the bands are computed as a function of a variety of tuning parameters. For AA systems, we only employ the LDA-SO functional, for reasons explained in Sec. \ref{section:Disc-of-relax}

\subsection{Electronic Structure}
\begin{figure}
\includegraphics[scale=.33]{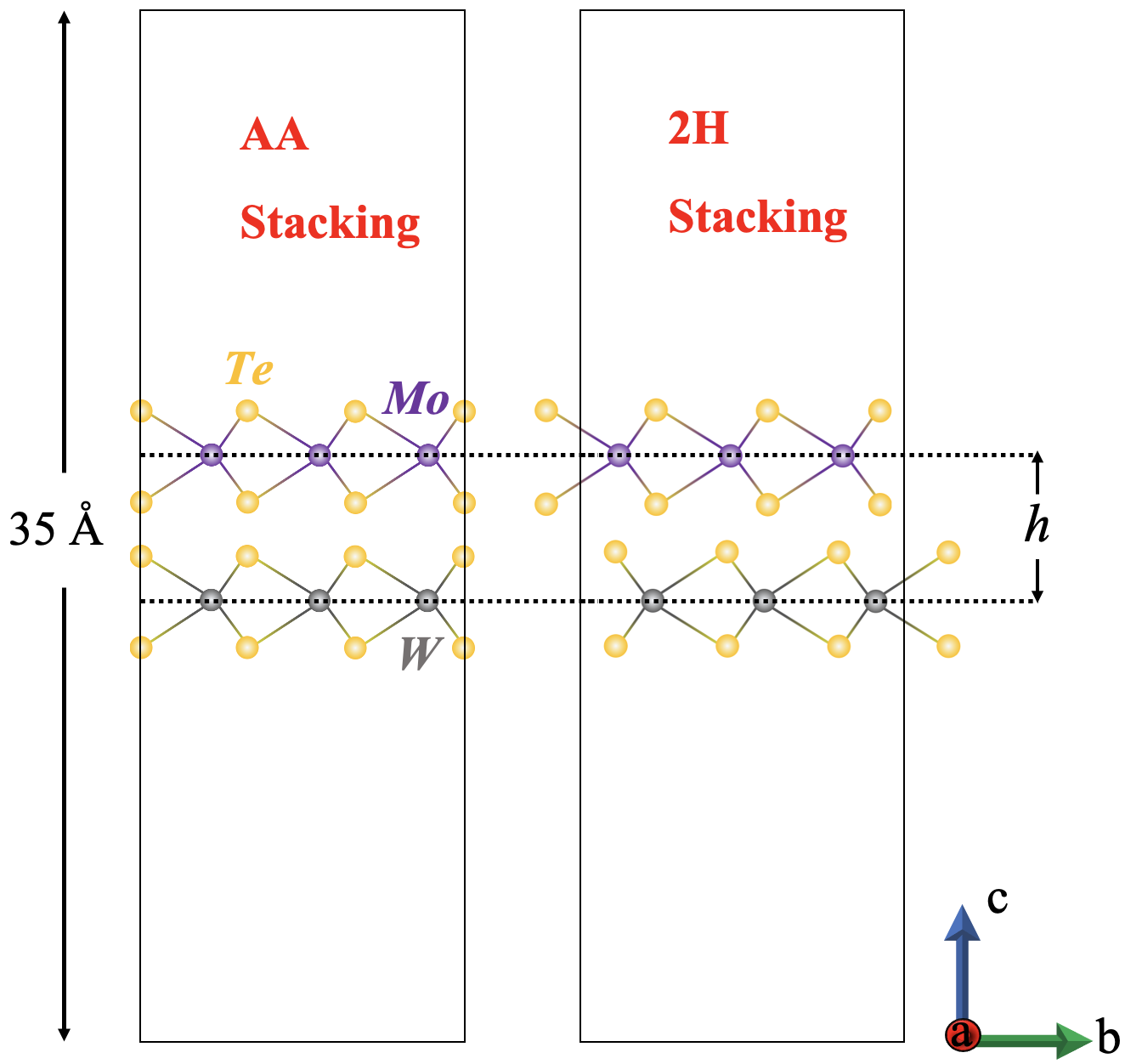} 
	\caption{ (color online). Example of AA and 2H stacking in the case of the free standing $\textrm{MoTe}_\textrm{2}$/$\textrm{WTe}_\textrm{2}$ TMD heterobilayer. A vacuum of magnitude 20 $\angstrom$ is added to the bulk c-axis lattice constant. $h$ is the interlayer separation, characterized by the vertical distance between metal atoms. }
	\label{fig:stacking-def}
\end{figure}

In order to comprehensively explore the energy competition between 
$\Gamma$ and K valleys, we perform Density Functional Theory (DFT) calculations using the full-potential Linearized Augmented Plane Wave (LAPW) method as provided by the WIEN2k package \cite{6165be91106043cc80ee83af3375a9c3,ss-calc-dft}. In addition to the standard Local Density Approximation (LDA) functional, we employ the modified Becke-Johnson (mBJ) functional to investigate the effect of electronic correlation \cite{original-BJ-potential,mBJ-potential}. It has been shown that the mBJ functional gives very accurate bandgaps in many transition metal oxides\cite{mbjvo2-1,wahilanbo2} and semiconductors,\cite{Borlido2019} including VO$_2$\cite{paezvo2,singhvo2} and monolayer TMDs,\cite{Li-bandgrapscreen} with much less computational time than hybrid functional, or $GW$ methods. Furthermore, the Local mBJ  functional has been developed \cite{local-mBJ} for accurate prediction of band gaps in systems with vacuum space, 
and works best for our study of the free standing bilayer TMDs illustrated in Fig. \ref{fig:stacking-def}. The energy convergence was chosen to be 0.1 mRy while the charge convergence was set to 0.001 $e^{-}$. In all cases, we include spin-orbit interactions and employ an additional $p_{1/2}$ radial basis function, called the Relativistic Local Orbital (RLO) provided by WIEN2k, for the metal atoms to improve the basis functions, aiding in convergence. For the band structure, we use an in plane momentum $\vec{k}$-mesh of $12 \times 12 \times 1$, and trace a $\Gamma \rightarrow \textrm{K} \rightarrow \textrm{M} \rightarrow \Gamma$ path in the Brillouin-zone.

The interlayer separation, $h$, defined henceforth as the distance measured purely along the $\hat{z}$-axis between metal atoms in each layer (shown in Fig. \ref{fig:stacking-def}) has a strong influence on the bilayer tunneling and as such is a key physical parameter \cite{https://doi.org/10.1002/aenm.201700571} that strongly influences the 
energy competition between the K and $\Gamma$ valleys. We therefore vary this parameter for values in the neighborhood of the relaxed interlayer separation, $h_r$, and examine the valley competition for each material in the 2H configuration. The total length along $\hat{z}$ (vacuum + c lattice constant) is kept fixed while the interlayer separation $h$ is varied. The magnitude of the a and b lattice constants are fixed throughout all calculations. This approach keeps the total system + vacuum unit cell volume fixed throughout the calculations, ensuring a consistent total energy.

In addition to interlayer separation, which can be adjusted by applying pressure, applied gate voltages are an experimentally accessible tuning parameter that influences the K-$\Gamma$ competition. \cite{E-field-MoS2,PhysRevB.84.205325,PhysRevB.94.241303,doi:10.1021/nn403738b,10.1063/1.4892798,Zheng_2016,PhysRevX.13.031037}. To finalize our discussion on key factors contributing to TMD band engineering, we perform electronic structure calculations for these materials under various electric fields.

\section{Results}
\begin{figure}
\includegraphics[scale=0.32]{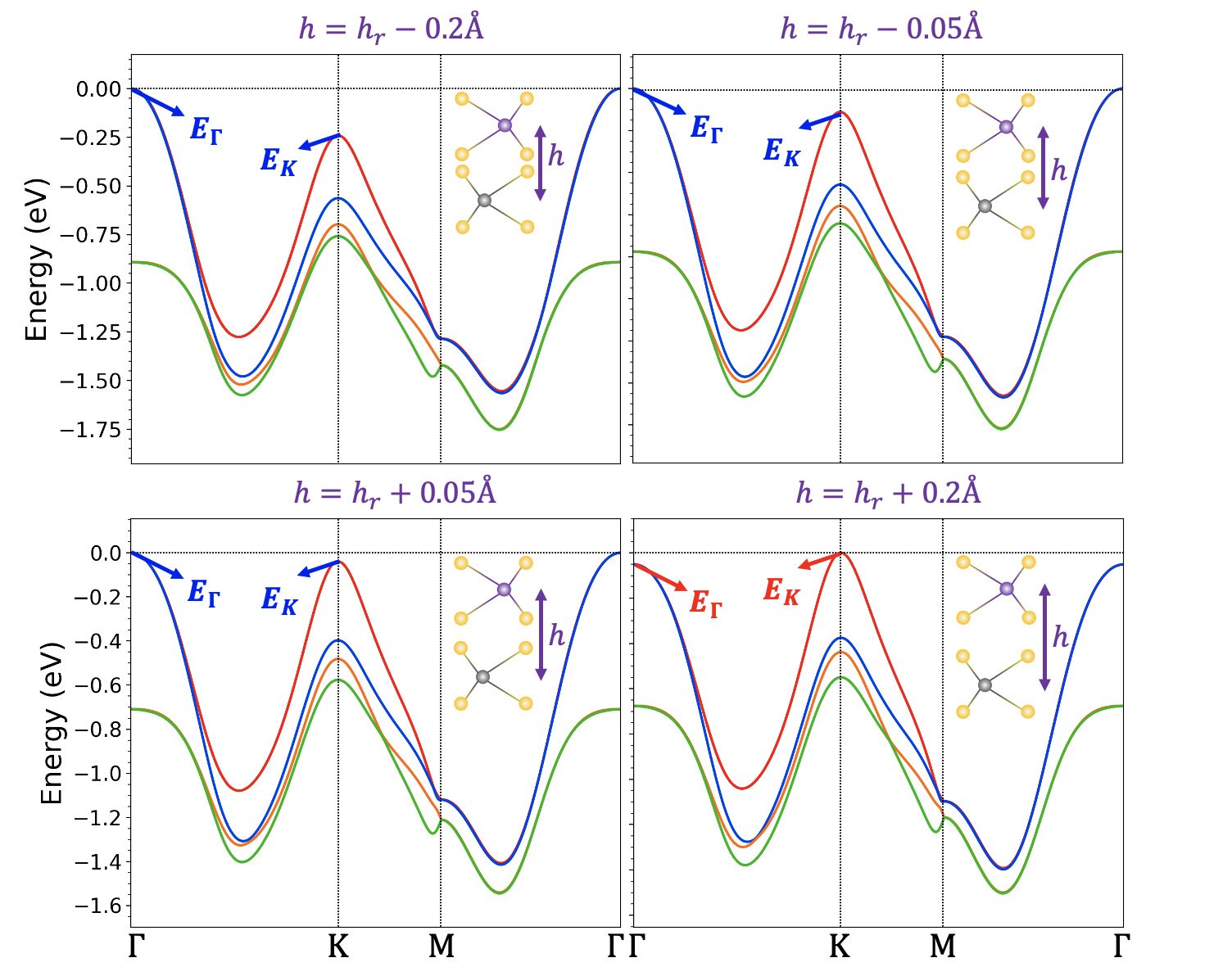} 
	\caption{ (color online). The top four valence bands computed with the LmBJ functional and spin-orbit coupling, as a function of $h$ for 2H-stacked $\textrm{MoS}_\textrm{2}$/$\textrm{WS}_\textrm{2}$ TMD bilayers. A sketch of each crystal structure is set into each plot. The energies of the  $\Gamma$ and K points are labelled as $E_\Gamma$ and $E_K$ respectively. These 
 results demonstrate that $E_K$ increases relative to $E_\Gamma$ as $h$ increases, as observed in all cases considered.}
    \label{fig:setup}
\end{figure}
We focus on exploring the key factors that determine the location of the valence band maximum for group VI TMD systems. As shown in Fig. \ref{fig:setup}, we denote the valence band energies at $\Gamma$ and K as $E_\Gamma$ and $E_K$, respectively and we employ DFT to calculate the energy difference $\Delta E_{K-\Gamma} = E_K -E_\Gamma$ as a function of interlayer distance $h$. To shed light on the role of electronic correlations in the $d$ orbitals of the transition metal atoms, we compare results obtained using the local mBJ functional with those obtained using the standard LDA functional.

\begin{figure}
\includegraphics[scale=0.42]{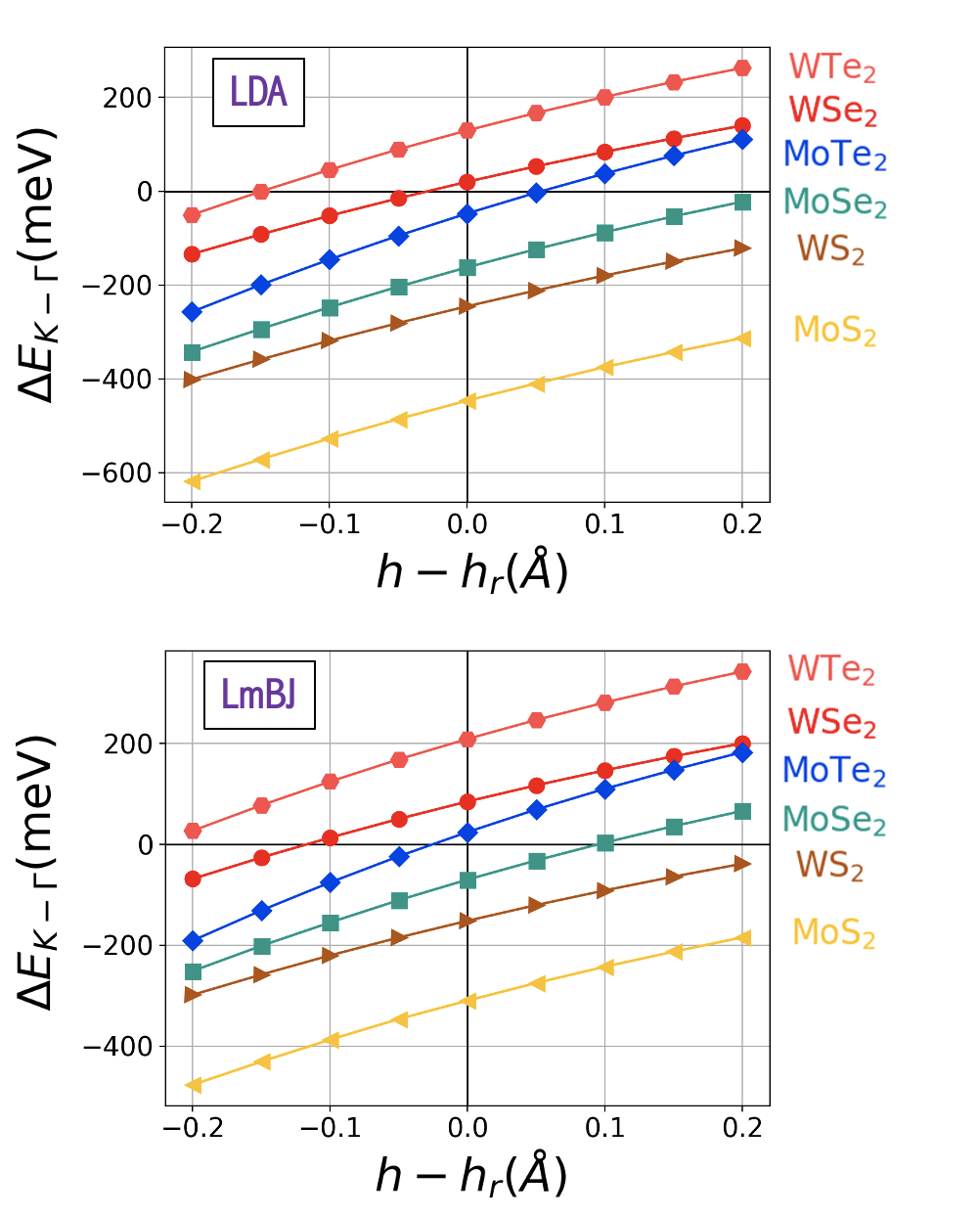} 
	\caption{ (color online). The energy difference between the local valence band maxima at $\Gamma$ and K points, $\Delta E_{K-\Gamma}\equiv E_K - E_\Gamma$, as a function of the interlayer distance $h$ calculated by LDA and local mBJ functionals for 2H stacked homobilayers. 
 The value of $h$ at which $\Delta E_{K-\Gamma}$ crosses from negative ($\Gamma$-valley) to positive (K-valley) decreases
 when correlations are included by using the local mBJ potential. All calculations include spin-orbit coupling.}
	\label{fig:homo-deltaE-2h}
\end{figure}
\begin{figure}
\includegraphics[scale=0.42]{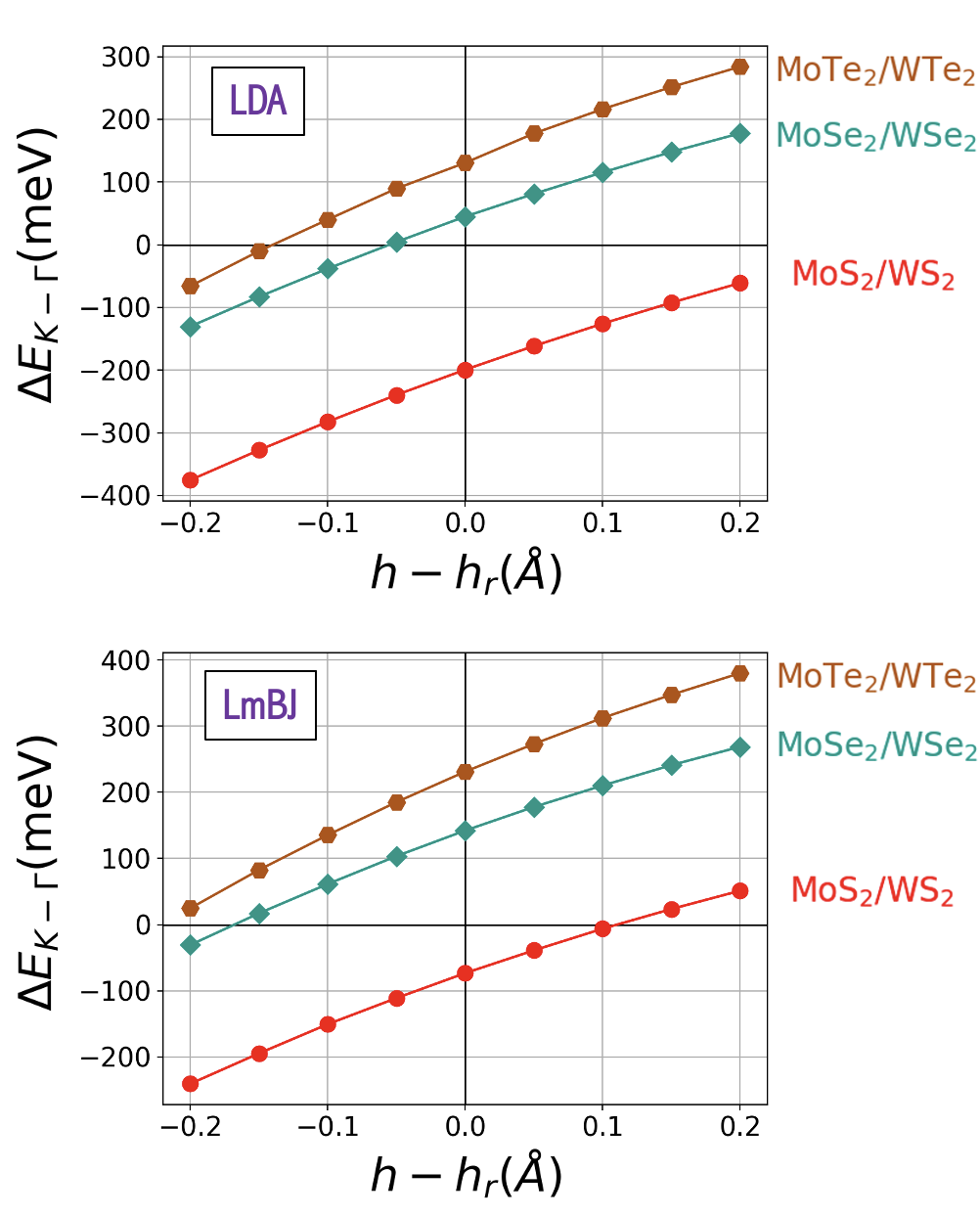} 
	\caption{ (color online). The energy difference between the local band maxima $\Gamma$ and K points $\Delta E\equiv E_K - E_\Gamma$ as a function of the interlayer distance $h$ calculated by LDA and local mBJ functionals for heterobilayers. Analysis of the band characteristics shows that the metal atom at the K valley of the highest valence band is that of Tungsten for all heterobilayers.}
	\label{fig:hetero-deltaE}
\end{figure}

Fig. \ref{fig:homo-deltaE-2h} summarizes our results for 2H stacked $\textrm{MX}_2$ homobilayers with $\textrm{M}=\textrm{Mo},\textrm{W}$ and $\textrm{X}=\textrm{S},\textrm{Se},\textrm{Te}$.
We plot the energy difference $\Delta E_{K-\Gamma}=E_K - E_\Gamma$ as a function of $h-h_r$ calculated by LDA (top) and local mBJ (bottom) functionals. Each system's equilibrium interlayer separation $h_r$ and its $\Delta E_{K-\Gamma}=E_K - E_\Gamma$ may be found in Table \ref{tab:all-summary}. Even though pressure cannot increase $h$, we include positive values of $h-h_r$ due to the range of possible separations reported in bulk TMDs. Here it can be observed that the use of the local mBJ potential tends to increase $\Delta E_{K-\Gamma}$. The same behavior is clearly observed in the case of heterobilayers as shown in Fig. \ref{fig:hetero-deltaE}. Selenide and Telluride heterostructures remain K-valley while shifting from equilibrium in nearly all cases. The sulfide heterostructure may achieve a K VBM at the highest separations, while neither sulfide-based homostructure can achieve a K VBM using a physically plausible increase in spacing alone.

After structural relaxation subsequent DFT calculations reveal that the total energy of each AA stacked system is consistently greater ($\sim 70-110$ meV per unit cell) than that of its corresponding 2H form, as may be seen in Table \ref{tab:all-AA-summary}. In this regard, the same overall trend as in the 2H case in the K-$\Gamma$ competition is expected as a function of $h-h_r$, but modulating the interlayer separation via pressure is not an experimentally feasible approach due to the predicted instability of the AA structure. The AA structures are only relaxed with LDA-SO, and the $\Delta E_{K-\Gamma}=E_K - E_\Gamma$ is computed for only the relaxed bilayer separation, as displayed in Table \ref{tab:all-AA-summary}.

\section{Discussion}

\subsection{Effects of element type}
\begin{figure}
\includegraphics[scale=.42]{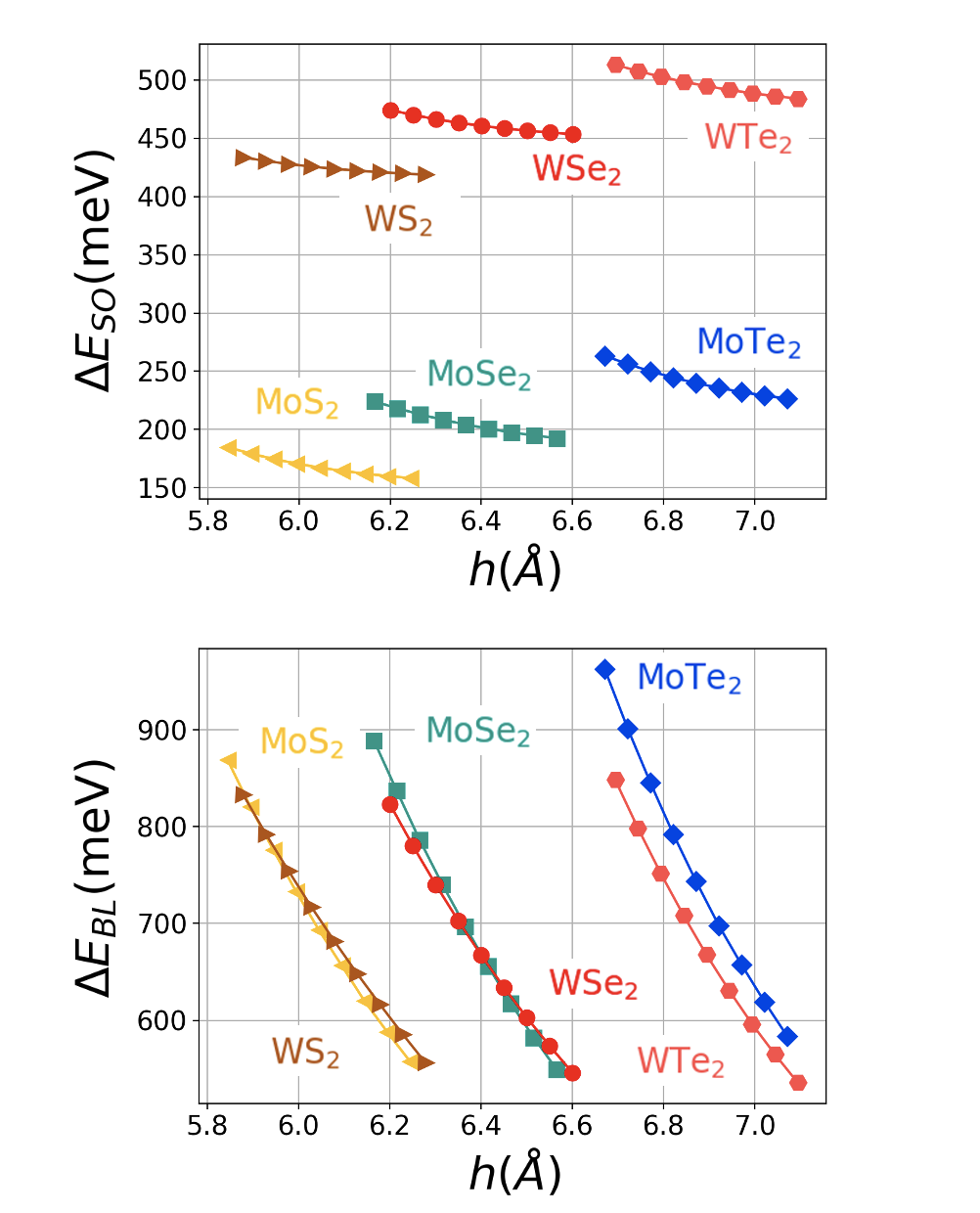} 
	\caption{ (color online). (top) The spin-orbit splitting $\Delta E_{SO}$ and (bottom) the bilayer splitting $\Delta E_{BL}$ as a function of the absolute interlayer separation $h$ for $\textrm{MoX}_2$ and $\textrm{WX}_2$ with $\textrm{X}=\textrm{S},\textrm{Se}$ and $\textrm{Te}$. The 2H stacked results using the local mBJ functional are presented.}
	\label{fig:trend}
\end{figure}

There are several element-related trends that can be seen in our results. To make the analysis concrete, we investigate the role of the element type in the 2H homobilayers only (Fig. \ref{fig:homo-deltaE-2h}). Firstly, we consider the trend of varying the metal atom in the $\textrm{MX}_2$ with the chalcogenide atom $\textrm{X}$ fixed. It can be seen clearly that in this case, $\textrm{WX}_2$ always has a larger value of $\Delta E_{K-\Gamma}$ than $\textrm{MoX}_2$ does. This can be attributed to the fact that the $\textrm{W}$ atom has a larger atomic number than the $\textrm{Mo}$ atom, and 
therefore larger spin-orbit coupling. To demonstrate this point, we define $\Delta E_{SO}$ as the energy difference between the top two valence states at the K point as shown in Fig. \ref{fig:setup}, which can serve as a good estimation of the spin-orbit coupling. Fig. \ref{fig:trend} (top) plots $\Delta E_{SO}$ as a function of the absolute interlayer separation $h$ for $\textrm{MoX}_2$ and $\textrm{WX}_2$ with $\textrm{X}=\textrm{S},\textrm{Se}$ and $\textrm{Te}$. Clearly, $\Delta E_{SO}$ depends strongly on the metal atom but weakly on the height $h$ and the chalcogenide atom $\textrm{X}$, which is expected since the spin orbit coupling is an intrinsic property of the metal atom.

Now we study the trend of varying the chalcogen atom in $\textrm{MX}_2$ with the metal atom $\textrm{M}$ fixed. We see a monotonic trend where $\Delta E_{K-\Gamma}$ is generally increased when going from $\textrm{S}\to \textrm{Se} \to \textrm{Te}$. We note that because chalcogenide atoms $\textrm{X}$ exist between metal atoms, as shown in Fig. \ref{fig:stacking-def}, the channels going through the $p$ orbitals of $\textrm{X}$ make the major contribution to the interlayer coupling. Generally speaking, as the atomic number of $\textrm{X}$ increases, orbitals on the outermost shell becomes more de-localized, which leads to a stronger wavefunction overlap with $d$ orbitals of the metal atom and consequently a larger bilayer splitting at the $\Gamma$ point. We define $\Delta E_{BL}$ as the energy difference between the top two valence states at $\Gamma$ point as shown in Fig. \ref{fig:setup}, which can be a good estimation of the interlayer coupling. $\Delta E_{BL}$ as a function of the absolute interlayer separation $h$ for $\textrm{MoX}_2$ and $\textrm{WX}_2$ with $\textrm{X}=\textrm{S},\textrm{Se}$ and $\textrm{Te}$ is presented in Fig. \ref{fig:trend} (bottom). It is clearly observed that bilayer splitting depends strongly on interlayer separation, and that the equilibrium separation of each material monotonically increases as $\textrm{S}\to \textrm{Se} \to \textrm{Te}$, indicating that stronger bilayer splitting heavily influences the relaxation.

\subsection{Discussion of Structural Relaxation}{\label{section:Disc-of-relax}}

\begin{figure}
\includegraphics[scale=.42]{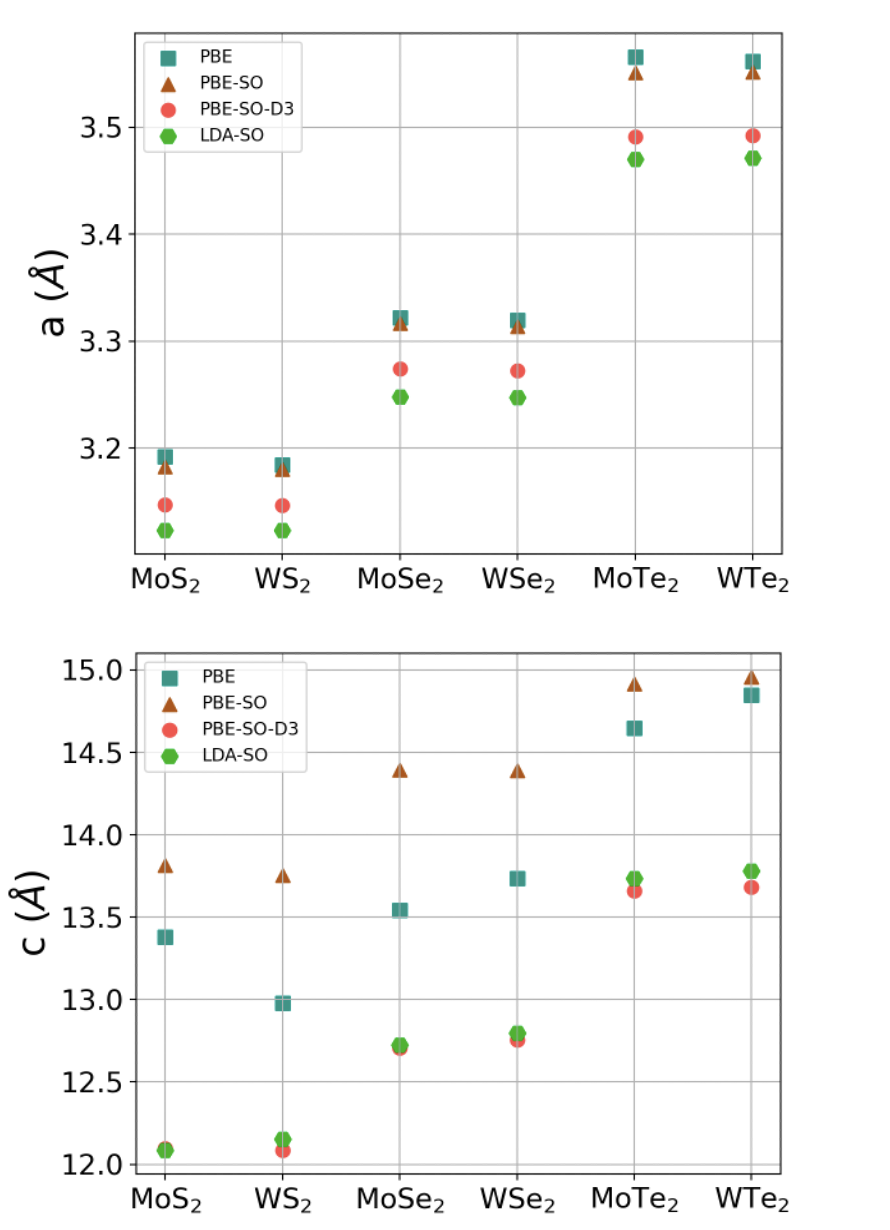} 
	\caption{ (color online). The (top) a and (bottom) c bulk lattice constants predicted by various potentials employed in VASP.}
	\label{fig:plot-relax}
\end{figure}

\begin{table*} 
 \begin{tabularx}{\textwidth}{X*{8}{>{\centering\arraybackslash}X}} 
 \textbf{2H}  [\textbf{\AA,meV}] & \color{red} bulk PBE \color{black}\textbf{(a/c)} & \color{red} bulk PBE-SO \color{black}\textbf{(a/c)} & \color{red} bulk PBE-SO-D3 \color{black}\textbf{(a/c)} &
 \color{red} bulk LDA-SO \color{black}\textbf{(a/c)} & \color{red} LDA-SO rel. \color{black}bulk $\textbf{h}$ $(\frac{c}{2})$ & \color{red} LDA-SO rel. \color{black} bilayer $\textbf{h}_\textbf{r}$ & \color{red} LDA+SO \color{black} K-$\Gamma$ & \color{red} Local mBJ \color{black} K-$\Gamma$\\ 
 \hline
 & & & & & & & & \\
 \color{blue} $\textrm{WS}_\textrm{2}$ & 3.184223 & 3.182996 & 3.146435 & 3.123324 & 6.076231 & 6.078068 & -245.48 & \textcolor{blue}{-151.53} \\
 & 13.37829 & 13.81513 & 12.08426 & 12.15246 &  &  &  &  \\
 & & & & & & & & \\
 \hline
 & & & & & & & & \\
 \color{blue} $\textrm{MoS}_\textrm{2}$  & 3.192238 & 3.182996 & 3.147325 & 3.123039 & 6.041991 & 6.044029 &  -446.68 & \textcolor{blue}{-309.88} \\
  & 13.37829 & 13.81513 & 12.09522 & 12.08398 & & &  &  \\
 & & & & & & & & \\
 \hline
 & & & & & & & & \\
 \color{purple} $\textrm{WSe}_\textrm{2}$ & 3.319933 & 3.314109 & 3.272225 & 3.247144 & 6.383032 & 6.400151 & 19.99 & \textcolor{purple}{85.02} \\
 & 13.73719 & 14.38835 & 12.75565 & 12.79507 & &  & & \\
  & & & & & & & & \\
  \hline
  & & & & & & & & \\
 \color{blue} $\textrm{MoSe}_\textrm{2}$  & 3.32226 & 3.316714 & 3.274179 & 3.247787 & 6.361821 & 6.364676 &  -162.43 & \textcolor{blue}{-69.83} \\
   & 13.54296 & 14.39224 & 12.70686 & 12.72364 & & &  &  \\
   & & & & & & & & \\
   \hline
   & & & & & & & & \\
 \color{purple} $\textrm{WTe}_\textrm{2}$  & 3.561547 & 3.551663 & 3.492066 & 3.471253 & 6.889546  & 6.893730 & 129.51 &  \textcolor{purple}{209.08} \\
   & 14.84723 & 14.95786 & 13.68179 & 13.77909 &  &  &  & \\
   & & & & & & & & \\
   \hline
   & & & & & & & & \\
 \color{purple} $\textrm{MoTe}_\textrm{2}$  & 3.565654 & 3.550858 & 3.490808 & 3.470279 & 6.866760 & 6.870946 &  -47.52 & \textcolor{purple}{24.39} \\
  & 14.64777 & 14.91474 & 13.65810 & 13.73352 &  &  &  &  \\
  & & & & & & & & \\
  \hline
  \hline
  & & & & & & & & \\
 \color{blue} $\textrm{MoS}_\textrm{2}/\textrm{WS}_2$  &  &  &  & 3.123039 &  & 6.049050 &  -199.32 & \textcolor{blue}{-73.35} \\
  &  &  &  &  &  &  &  &  \\
  & & & & & & & & \\
    \hline
  & & & & & & & & \\
 \color{purple} $\textrm{MoSe}_\textrm{2}/\textrm{WSe}_2$  &  &  &  & 3.247787 &  & 6.379695 &  44.96 & \textcolor{purple}{142.17} \\
  &  &  &  &  &  &  &  &  \\
  & & & & & & & & \\
    \hline
  & & & & & & & & \\
 \color{purple} $\textrm{MoTe}_\textrm{2}/\textrm{WTe}_2$  &  &  & & 3.470279 &  & 6.879600 & 130.65 & \textcolor{purple}{231.21} \\
  &  &  &  &  &  &  &  &  \\
  & & & & & & & & \\
  \hline
\end{tabularx} 
\caption{\label{tab:all-summary} Relaxed bulk and bilayer structural and electronic data summarized in $\angstrom$ and $meV$. Materials highlighted in blue are $\Gamma$ valleys while those in purple are K valleys according the LmBJ.}
\end{table*}
The structural relaxation is performed in VASP, following the procedure outlined in Sec. \ref{section:Struct-relax}. The resulting structural data is summarized in Table \ref{tab:all-summary}. For clarity and discussion, we additionally include a plot of the bulk a and c lattice constants in Fig. \ref{fig:plot-relax}. There are a number conclusions that may be drawn from the relaxation procedure. Inclusion of the vacuum after LDA-SO bulk relaxation can be seen to increase the metal-to-metal atom distance in bilayers, which will reduce interlayer tunneling. 
The reduction of $h$ in bulk systems is expected since the van der Waals attraction between layers is expected to be stronger when every layer has two neighboring layers. If this consideration is correct, then the encapsulating layers present in most experimental systems may also play a role in determining the equilibrium layer separation. Our DFT calculations confirm this assertion as we see a boost in the K valley while comparing bulk to bilayer results. As we employ both VASP and WIEN2k, we may also comment that the LDA+SO band structures of VASP's pseudo-potential method and the all-electron method of WIEN2k are nearly identical if the same structure is provided. \footnote{Furthermore, we find that the total energy of the bilayers predicted by LDA+SO using WIEN2k is \textit{in general} at the global minimum when the equilibrium interlayer separation is used. The energy generally grows as we consider the shifting of $\pm 0.2$ \AA, with a sharper growth with positive shifting in interlayer separation.}

\begin{table}
 \begin{tabular}{c|p{1.5cm}|p{1.4cm}|p{1.45cm}|p{1.3cm}}
 \textbf{AA} & \color{red} LDA+SO \color{black} K-$\Gamma$ (meV) & \color{red} Local mBJ \color{black} K-$\Gamma$ (meV) & $\Delta E_{AA-2H}$ via LDA+SO (Ry) & h (\AA)\\ 
 \hline\hline
 $\textrm{WS}_\textrm{2}$ & 87.4 & 126.8 & 5.69 & 6.776537\\
 \hline
 $\textrm{MoS}_\textrm{2}$  & -86.1 & 3.0 & 5.36 & 6.769552\\
 \hline
  $\textrm{WSe}_\textrm{2}$  & 368.3 & 399.1 & 5.72 & 7.128487\\
 \hline
 $\textrm{MoSe}_\textrm{2}$ & 218.2 & 271.9 & 6.34 & 7.100237\\
 \hline
  $\textrm{WTe}_\textrm{2}$ & 559.3 & 610.3 & 8.09 & 7.769766\\
   \hline
  $\textrm{MoTe}_\textrm{2}$  & 443.3 & 483.0 & 8.19 & 7.761531\\
  \hline
 $\textrm{MoS}_\textrm{2}/\textrm{WS}_2$  & 147.0 & 215.0 & 5.63 & 6.769726\\
    \hline
 $\textrm{MoSe}_\textrm{2}/\textrm{WSe}_2$  & 406.2 & 453.2 & 6.12 & 7.111891\\
    \hline
 $\textrm{MoTe}_\textrm{2}/\textrm{WTe}_2$  & 588.8 & 649.9 & 8.10 & 7.767940\\
\end{tabular} 
\caption{\label{tab:all-AA-summary} The K-$\Gamma$ competition for relaxed AA structures, obtained using just LDA+SO and the Local mBJ functionals, in \AA, meV for the K-$\Gamma$ and Rydberg for $\Delta E_{AA-2H}$. The quantity $\Delta E_{AA-2H}=E^{LDA+SO}_{AA}-E^{LDA+SO}_{2H}$, where $E^{LDA+SO}_{AA}$ and $E^{LDA+SO}_{2H}$ are the total energy in Rydberg for LDA+SO of the AA structure and 2H structures, respectively. The difference is always positive due to the sharp raise in total energy of the AA structures, observed in relaxation.}
\end{table}

It is clear from the step pattern in Fig. \ref{fig:plot-relax} that the magnitude of the lattice constants is determined mainly by the type of chalcogen in the structure. This is particularly true for the methods of PBE-SO-D3 and LDA-SO in the case of the c lattice constant. With comparison to the range of literature reported bulk values \cite{Puotinen:a03172,PhysRevB.35.6195,PhysRevB.12.659,SCHUTTE1987207,Molina_S_nchez_2011,crystal-chem-of-tmds}, we conclude that PBE and PBE-SO both tend to overestimate the lattice constants alone, and the inclusion of van der Waals is necessary. As expected \cite{Molina_S_nchez_2011}, even in the absence of the van der Waals force LDA-SO agrees with PBE-SO-D3 quite well for layered systems such as the group VI TMDs, and is in some cases more accurate with respect to experiments. Hence, we employ only LDA-SO for the AA stacking. Our bilayer electronic structure calculations are carried out using constants from the LDA-SO approach, as explained in \ref{section:Struct-relax}

\subsection{General discussion}
The energy competition between $\Gamma$ and K valleys is influenced by the strengths of interlayer tunneling, spin orbit coupling, and electronic correlations. Generally speaking, interlayer tunneling produces a more sizable bilayer splitting at the $\Gamma$ point pushing the top valence state up, with negligible effect at the K point. As a result, a smaller distance between the bilayers favors the valence band maxima to be at $\Gamma$ point due to the larger bilayer splitting. For heterobilayers, the symmetry between layers is broken and the bilayer splitting at the K is not negligible, which explains why the heterobilayers tend to favor the K valley when compared to the homobilayers of their parent atoms. On the other hand, the energy splitting due to spin-orbit coupling is very small near the $\Gamma$ point but much more pronounced near the K point. Consequently, stronger spin-orbit coupling favors the top valence band maxima to be at K point. Since varying the distance between bilayers can modulate the interlayer tunneling strength, without influencing the spin orbit coupling strongly (see Fig. \ref{fig:trend}), it is an ideal experimental parameter to engineer the $\Gamma$-K competition.

Because the highest energy valence bands are composed primarily of $d$-orbitals from the transition metal atoms, electronic correlations also influence the $\Gamma$-K energy competition. Given that the wave function is more extended for states near the $\Gamma$ point and more localized for states near the K point, correlations tend to push states near the K point to higher relative energy. This trend is observed in all of our DFT calculations. One may observe that in Molybdenum-based structures, the inclusion of electronic correlation produces a larger upward push, due to the 4$d$ orbital characteristic as opposed to the 5$d$ of the stable Tungsten-based structures. 

With respect to the stacking order, we note that the AA stacked materials possess a much larger metal to metal distance when fully relaxed compared to their 2H counterparts. As a result, we expect that the AA bilayer splitting is smaller than in 2H, contributing to a lower $\Gamma$ valley.
In Table \ref{tab:all-AA-summary}, we see that all of the equilibrium AA stacked materials are K-valleys for LmBJ, due primarily to this weaker interlayer tunneling. As in the 2H cases, the correlation effect of the LmBJ potential is to push the K-valley up, but the magnitude of the increase is lesser in AA materials than in 2H.

\section{External tuning parameters}

\subsection{Methods to change interlayer distance}
Pressure engineering has been widely used as one of the ways to tune an interlayer distance in bilayer and multilayer TMDs \cite{bhatt2012,fan2015,dou2014,su2014,xia2021,nayak2014,nayak2015}. Decrease in the interlayer distance has been achieved experimentally by applying hydrostatic pressure through a DAC (Diamond anvil cell) \cite{nayak2014,nayak2015,dou2014,xia2021}. Pressure used to decrease the interlayer separation can be used to modify the band-gap, and in the semiconducting regime the change of interlayer distance estimated from optical probes can be up to 10\% within 10 GPa. \cite{nayak2015,nayak2014,xia2021}.
While we do not attempt to make a quantitative study of the pressure-induced effects, we would like to see if our results are reasonably consistent with experiments. For this purpose, we employ the formula used in Ref. [\onlinecite{Bhattacharyya2012}] that approximates pressure as 
\be
P \approx \frac{E_{tot}(h)-E_{tot}(h_r)}{A(h_r-h)}
\ee
where $E_{tot}(h)$ is the total energy (from LDA+SO) at interlayer separation $h$, $h_r$ is the relaxed interlayer separation, and $A$ is the unit cell area in the $ab$ plane. For the heterobilayer MoSe$_2$/WSe$_2$ that has been experimentally studied\cite{xia2021}, we find that in order to achieve 3\% change of the height, the required pressure is predicted to be $\sim 1$ GPa. Although the order of magnitude seems to be reasonably within the experimental range, more calculations must be done to have a quantitative description of the pressure effects and more experimental data will be needed for comparison. Applying pressure that decreases the layer separation can only enable transitions from K to $\Gamma$ - and not the reverse transition.

While applying pressure is an experimentally feasible way to decrease the interlayer distance, the strain engineering offers new possibilities to increase the interlayer distance. By applying tensile or compressive strain to the bilayer TMD, interlayer distance can be either increased or decreased. Transition between direct and indirect bandgaps has been reported in experimental and theoretical studies.\cite{Kumar_2013,pak2017}

\subsection{Electric gating}
\begin{figure}
\includegraphics[scale=0.245]{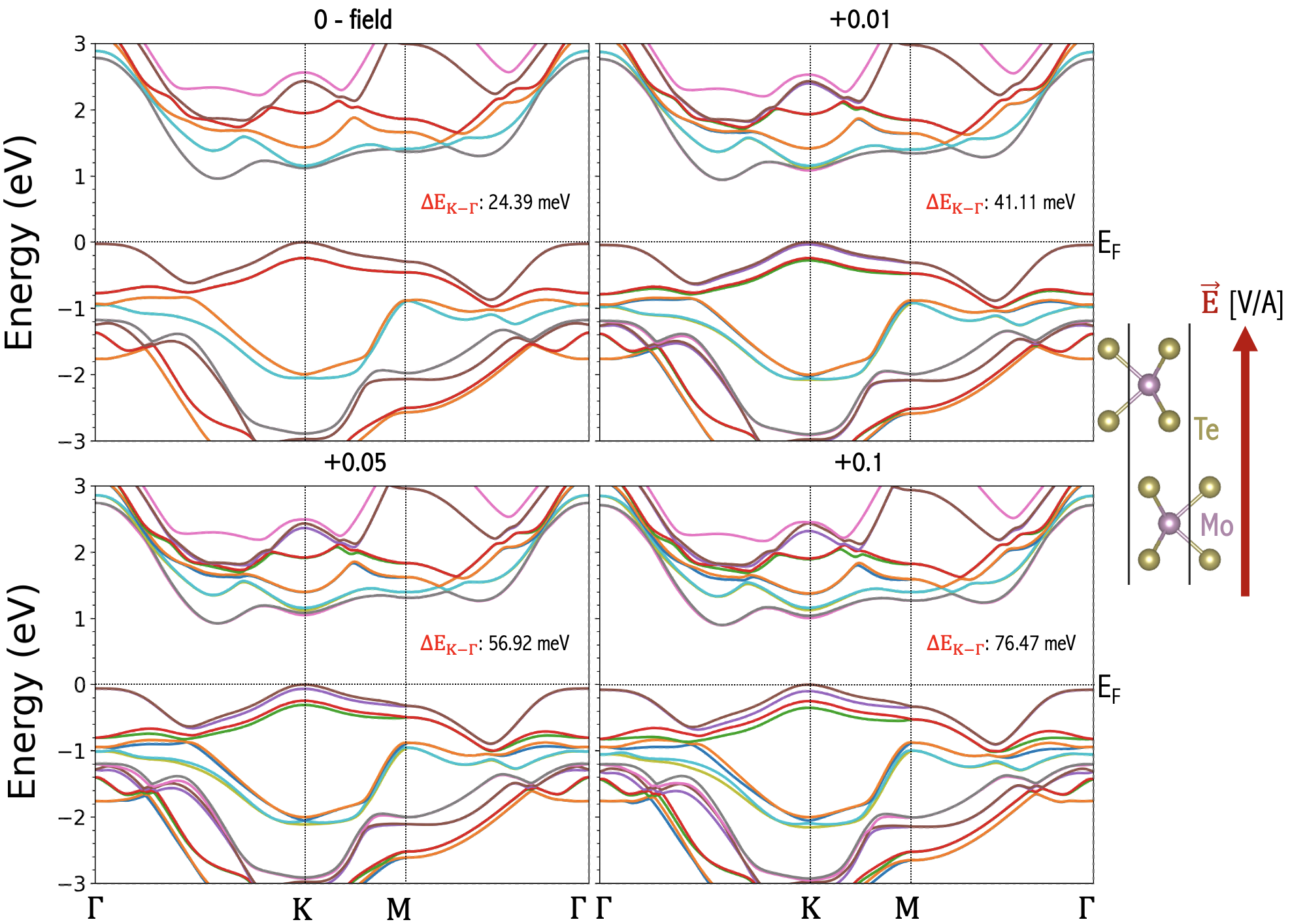} 
	\caption{ (color online). The electric field effect on the band structure of 2H $\textrm{MoTe}_2$ bilayers. For homobilayers the direction of the field is not relevant. Starting from 0 applied field at the top left, the magnitude of the electric field is increased yielding an enhanced splitting near the K valley, and in between the $\Gamma$ and K valleys, without strongly affecting the $\Gamma$ valley. }
	\label{fig:homo-efield}
\end{figure}

\begin{figure}
\includegraphics[scale=0.245]{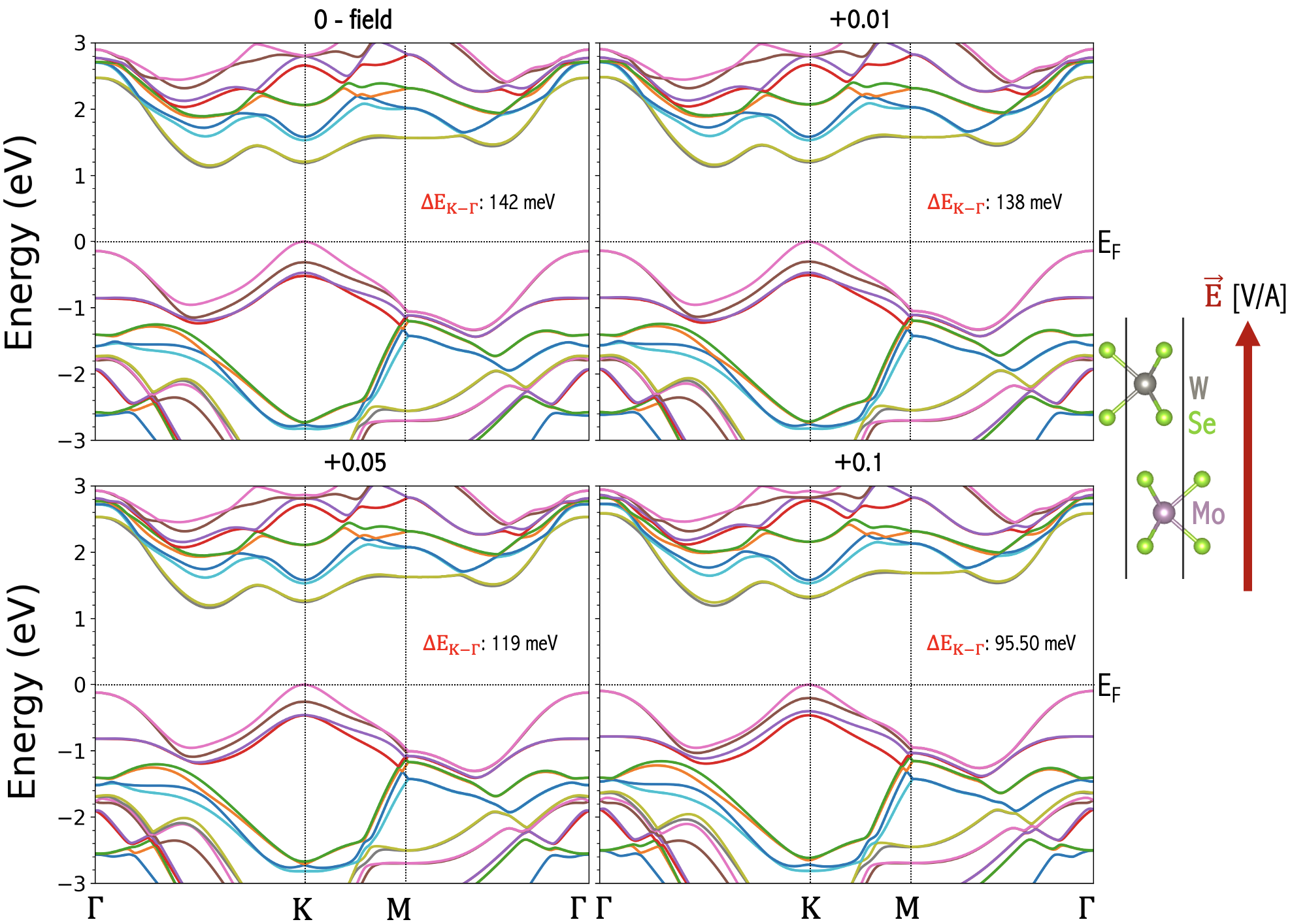} 
	\caption{ (color online). The electric field effect on the band structure of 2H $\textrm{MoSe}_2/\textrm{WSe}_2$ bilayers. Starting from 0 applied field at the top left, the field pointing from the bottom layer of $\textrm{MoSe}_2$ to the top layer of $\textrm{WSe}_2$ is increased in magnitude. For heterobilayers, we find that the direction is important (see text).}
	\label{fig:hetero-efield-plus}
\end{figure}

\begin{figure}
\includegraphics[scale=0.245]{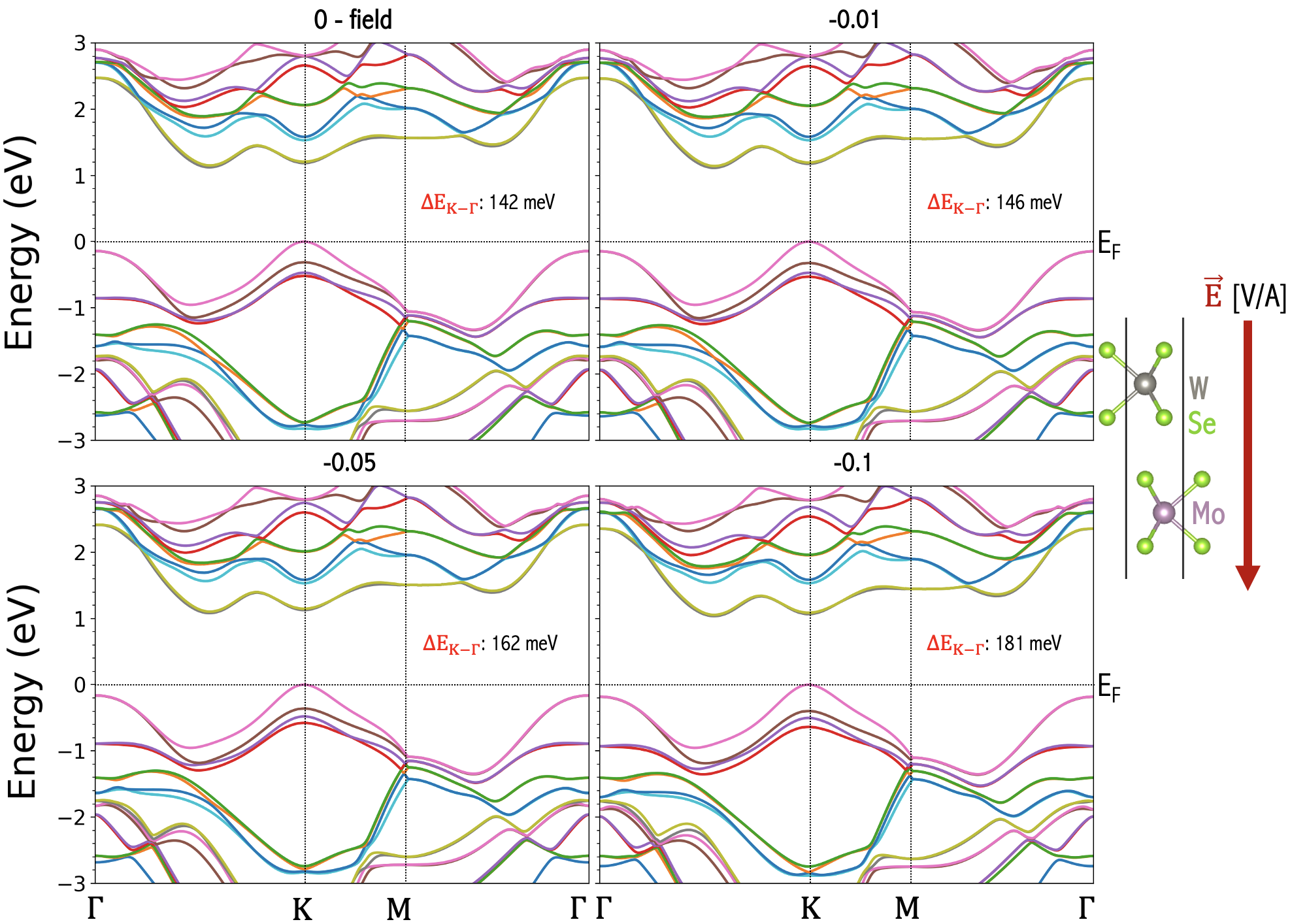} 
	\caption{ (color online). The electric field effect on the band structure of 2H $\textrm{MoSe}_2/\textrm{WSe}_2$ bilayers. Starting from 0 applied field at the top left, the field pointing from the top layer of $\textrm{WSe}_2$ to the bottom layer of $\textrm{MoSe}_2$ is increased in magnitude. For heterobilayers, we find that the direction is important (see text).}
	\label{fig:hetero-efield-minus}
\end{figure}

The most convenient way of engineering bilayer electronic structure is to apply electrical gate fields. \cite{E-field-MoS2,PhysRevB.84.205325,PhysRevB.94.241303,doi:10.1021/nn403738b,10.1063/1.4892798,Zheng_2016}. We employ the "zig-zag" approach in WIEN2k \cite{wien2k,wien2k2020} to apply a simulated gate voltage across the layers of $\sim 0.2$V, yielding transverse electric fields with magnitudes of $\sim 0.01-0.1$ V/A in the samples. Examples of the resulting band structure change for the experimentally relevant \cite{Cai_2023,Mak2018,zeng2023integer,Jin_2021,doi:10.1126/science.aac7820} cases of homobilayer $\textrm{MoTe}_2$, and heterobilayer of $\textrm{MoSe}_2/\textrm{WSe}_2$ are shown in Fig. \ref{fig:homo-efield}, and Figs. \ref{fig:hetero-efield-plus}, \ref{fig:hetero-efield-minus}, respectively. We note that the electric field produces the stronger effect near K than $\Gamma$ valleys. That the K valley is influenced more than the $\Gamma$ valley by the external field can be explained by the same model presented in the study of the bilayer graphene\cite{PhysRevB.75.155115}. In essence, since the external electric field produces an electric potential difference between layers, its effect can be described by a two-band model capturing the layer degrees of freedom in Equation 5 of Ref. [\onlinecite{PhysRevB.75.155115}]. Firstly, in the two-band model, the electric potential difference appears in the diagonal terms, whose effect in the eigenenergies will be reduced in general in the presence of the off-diagonal terms due to the interlayer tunneling. As a result, the shift of the eigenenergies is expected to be larger at momentum where tunneling is small. This explains our observation of the external electrical field having stronger effect near K than $\Gamma$ valleys. We note that the evolution of the $\Delta E_{K-\Gamma}$ is further complicated by the screening effect due to the Hartree potential that is shown to be large and well-captured by DFT calculations. In general, the screening effect decreases the electric potential difference actually seen in the bilayer systems in a non-monotonic way. 

Another intriguing observation is that for homobilayers (Fig. \ref{fig:homo-efield}) direction of the field does not matter as both layers maintain the same on-site potential in each layer and have a mirror symmetry between layers, and the field makes a positive contribution to $\Delta E_{K-\Gamma}$. For heterobilayers (Figs. \ref{fig:hetero-efield-plus}, \ref{fig:hetero-efield-minus}), we observe that changing the field direction produces opposite effects. If the field points from $\textrm{MoSe}_2$ to $\textrm{WSe}_2$, the top valence band (which is dominated by d orbitals of W atom) is pushed away from the Fermi energy and the gap is enlarged resulting in a reduction in $\Delta E_{K-\Gamma}$. If the field points from $\textrm{WSe}_2$ to $\textrm{MoSe}_2$, the top valence band is pushed closer to the Fermi energy and the gap is reduced. This can also been explained by the two-band model\cite{PhysRevB.75.155115}. Because in the heterobilayer the on-site energy is already different between layers, a diagonal term in the two band model breaking the layer symmetry is already present at zero electric field. Therefore, the external field can drive K closer to or further away from the VBM in heterobilayers, depending on whether the direction of the electric field aligns with the chemical potential difference. Accordingly, the gate voltage is a powerful tuning parameter in TMD systems.

\section{Conclusion}
In this paper, we have explored fundamental ingredients that influence the competition between valleys within the valence bands of the experimentally promising group VI transition metal dichalcogenide systems. Our results show that several physical ingredients, namely the interlayer tunneling, spin-orbital coupling, gate voltage, and the electron correlation between the $d$-orbitals of the metal atoms, all play a significant role in the calculated energy at the K and $\Gamma$ points particularly at the top of the valence band. 

In order to study these physical processes, we first completed comprehensive structural relaxation for 2H and AA stacking configurations using VASP. Our results confirm that for the PBE functional, both spin orbit and van der Waals corrections (SO-D3) are necessary to ensure that lattice constants aren't overestimated, which would in turn misrepresent the bilayer splitting at $\Gamma$. We find that the local density approximation with spin orbit correction agrees well with the full PBE-SO-D3, which provides a computationally practical alternative for structural calculations in group VI TMDs. Our findings lead us to the conclusion that the magnitude of lattice constants is primarily determined by the chalcogen in the system. We also report that for the LDA+SO band structure, the pseudo-potential and all-electron approaches of VASP and WIEN2k, respectively, strongly agree. 

Using density function theory, we performed a systematic study of the interlayer splitting as a function of layer separation, and found that the energy at the K point can be driven higher as a result of increasing separation, while conversely, the energy at $\Gamma$ is decreased. Furthermore, we applied an external field and observed an increase in the K valley energy in homostructures, and a direction-dependent movement of the same valley in heterostructures. Our DFT calculations employed both the standard LDA function and the Local mBJ functional. As the $d$-orbitals of the transition metals from the parent layers are known to be dominant contributors to the top valence bands, including the electronic correlation effect from the mBJ potential is physically warranted. We find that the inclusion of the electronic correlation further pushes the K point nearer to the Fermi energy, with a negligible effect on the $\Gamma$ point, consistent with the idea that the wave function is more localized at K. Additionally, we find that the spin degeneracy at K is lifted with the inclusion of the spin-orbit coupling, further separating the top valence bands. At the same time, this inclusion does not drastically change the energy at the $\Gamma$ point.

Our findings show that applying gate voltage and tuning of the interlayer separation are powerful tuning knobs for topological and band engineering. For example, K and $\Gamma$ valleys may be tuned to be nearly degenerate within realistic range for experiments, which can lead to an effective two-dimensional strongly correlated two-orbital material after an induced moir\'{e} twist. The inclusion of the electronic correlation causes this K-point energy to raise even higher, making a crossover separation more achievable. Increasing the separation drives the K point higher \textit{without} affecting the degeneracy-lifting spin-orbit interaction, meaning that our findings outline an experimentally feasible approach for TMD engineering. 

\section{Acknowledgement}
S. O., E.J., and W.-C.L. were supported by the Air Force Office of Scientific Research
Multi-Disciplinary Research Initiative (MURI) entitled, “Cross-disciplinary
Electronic-ionic Research Enabling Biologically Realistic Autonomous Learning
(CEREBRAL)” under Award No. FA9550-18-1-0024 administered by Dr. Ali Sayir. 
A.H.M was supported by U.S. Department of Energy, Office of Basic Energy Sciences (DE-SC0021984).

\bibliography{TMD-GK}
\end{document}